\documentclass{elsart}
\usepackage{psfig}
\usepackage{wrapfig}

\newcommand{\CO}{$\rm { }^{57}Co$}

\newcommand{\SE}{\rm steering electrode}

\newcommand{\PE}{photoelectron}
\newcommand{\D}{diffusion}
\begin{document}
\begin{frontmatter}
\title{Investigation of charge sharing among electrode strips for a CdZnTe 
detector}
\author[CASS]{E. Kalemci\thanksref{CA}}
\author[CASS]{ J. L. Matteson}
\thanks[CA]{Corresponding author. e-mail: emrahk@mamacass.ucsd.edu\\
Tel: 858 534 64 31}
\address[CASS]{Center for Astrophysics and Space Sciences, University of California 
\\ San Diego
\\9500 Gilman Dr., La Jolla, CA 92093-0424\\}

%


\begin{abstract}
We have investigated charge sharing among the anode strips of a CdZnTe (CZT) 
detector using a $\rm 30\,\mu m$ collimated gamma-ray beam. We compared the 
laboratory measurements  with the predictions from our modeling of the charge 
transport within the detector. The results indicate that charge sharing is a 
function of the interaction depth and the energy of the incoming photon. Also, 
depending on depth, a fraction of the electrons might drift to the inter-anode
 region causing incomplete charge collection. Here, we show that \PE\ range and
 \D\ of the charge cloud are the principal causes of charge sharing and obtain
 limits on the size of the electron cloud as a function of position in the 
detector.
\end{abstract}
\end{frontmatter}
\section{Introduction}
 
 CZT has desirable features for detection of high energy X-rays and low energy 
gamma-rays. Due to its high atomic number, Z $\rm \sim$50, photoelectric 
interactions dominate up to 250 keV. Its large bandgap ($\rm \sim$1.54 eV) 
and high bulk resistivity ($\rm \sim$$\rm 10^{11}\,\Omega\,cm$) result in low 
leakage current and noise\cite{Kalemci99}, making it possible to achieve good
energy resolution at room temperature. Its energy response is linear over the 
energy range of five to several hundred keV\cite{SPIE2859}. However, with a
conventional planar configuration, a low energy tail appears due to incomplete 
hole collection in the pulse height spectrum of a monoenergetic source. Strip 
or pixel detectors can eliminate this problem by virtue of the small pixel 
effect\cite{Barret95,Luke95}. 

  We have developed and tested position-sensitive cross-strip CZT detectors 
for use in X-ray astronomy in collaboration with Washington University in St. 
Louis\cite{Kalemci99,SPIE2859,SPIE3445}.  Previously, we modeled charge
 drift in the detectors and charge induction on the electrodes as a function 
of time to obtain a better understanding of these detectors. The model agreed 
very well with the measurements of total charge. The details of this modeling 
can be found in Kalemci et al.\cite{Kalemci99}.
 
  The motivations behind the present study were to test the part of the 
simulation dealing with the charge drift trajectories and to understand how 
charge sharing among the anodes affects the performance of our detector. In 
this paper, the term ``charge sharing'' means that the electrons created by 
some interactions are collected by more than one anode strip.  
We attribute charge sharing to \PE\ range and diffusion of the resulting 
electron cloud. Depending on the distance between electrodes and the potential 
distribution in the detector, some of the electrons in this cloud might drift 
to inter-anode regions and cause incomplete charge collection, thereby 
distorting spectra and reducing energy resolution. Moreover, the size of the 
charge sharing region may limit the achievable spatial resolution. In our 
study, we measured the size of the \D\ cloud and its dependence on depth of 
interaction.

  Our experimental approach was to use a collimated gamma-ray beam to scan across
the electrodes of a cross-strip CZT detector and to analyze the observed pulse
 heights. Similar studies have been done by other groups with pixel detectors. 
Du et al.\cite{Du99} and He et al.\cite{He00} investigated charge charing in 
3-D position sensitive CdZnTe spectrometers and obtained sizes of electron 
clouds for different gamma-ray sources. Bolotnikov et al.\cite{Bolotnikov99} 
studied the charge loss to inter-pixel gaps. Prettyman et al.\cite{Prettyman99}
 examined the charge sharing effect on coplanar grid detectors and 
investigated the gap events.
\section{CZT strip detector}

\begin{figure}[tb]
\centering\psfig{figure=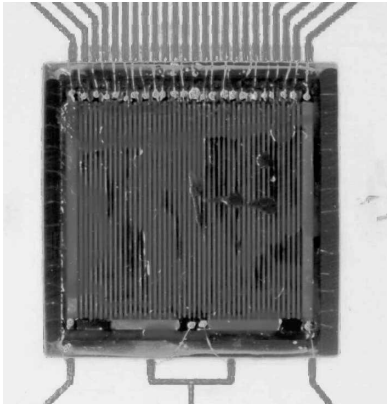}
\caption[det]
{\label{fig:det}
Prototype CZT strip detector. The anode side is shown, with its 22 anode 
strips with $\rm 500\,\mu m$  pitch and 22 interlaced steering electrodes. 
Total size is $\rm 12 \times 12 \times 2\, mm^{3}$. }
\end{figure}

\begin{figure}[tb]
\centering\psfig{figure=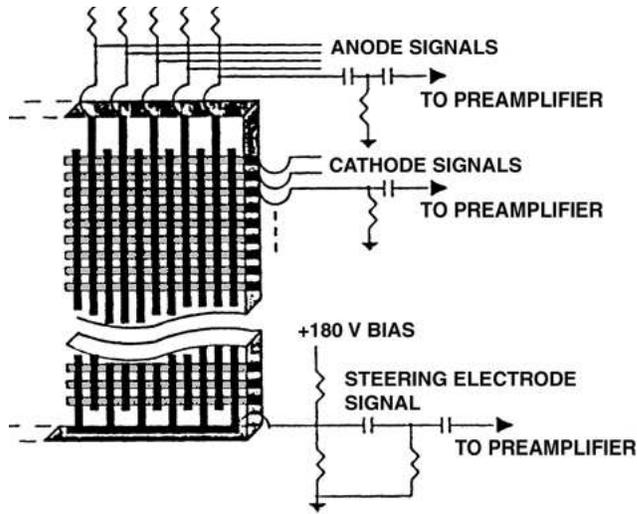,clip=,bbllx=0mm,bblly=0mm,bburx=85mm,bbury=70mm}
\caption[elc]
{\label{fig:elc}
Detector electrode and bias network connections.}
\end{figure}

  For our measurements, we used the UCSD-WU laboratory prototype detector (See 
Fig.~\ref{fig:det}) which was developed to study techniques for X-ray imaging 
applicable to astronomical instruments such as HEXIS\cite{SPIE3445} and 
MARGIE\cite{MARGIE}. This $\rm 12 \times 12 \times 2\,mm^{3}$ detector is 
configured with 22 anode strips on one side and 22 cathode strips on the other
 side. The cathode strips are orthogonal to the anode strips creating, in 
effect, a grid of $\rm 500\,\mu m$ pixels. To enhance charge collection on the 
anodes, a set of steering electrodes is interlaced with the anodes. These \SE 
s are all connected to each other (see Fig.~\ref{fig:elc}). Anodes and \SE s 
are $\rm 100\;\mu$m wide, while cathodes are $450\;\mu$m wide. The pitch size 
is $\rm 500\;\mu$m. The anodes are biased at 200 V and the \SE\ at 180 V, 
causing most electrons drifting towards the anode side of the detector to be 
directed away from the inter-anode region and towards an anode. A ceramic 
carrier holds the detector, that is always illuminated from the cathode side 
to minimize signal loss due to hole trapping. It was manufactured from 
``discriminator grade'' material by eV Products.
\section{Charge Collection Model}

\begin{figure}[t]
\centerline{\psfig{figure=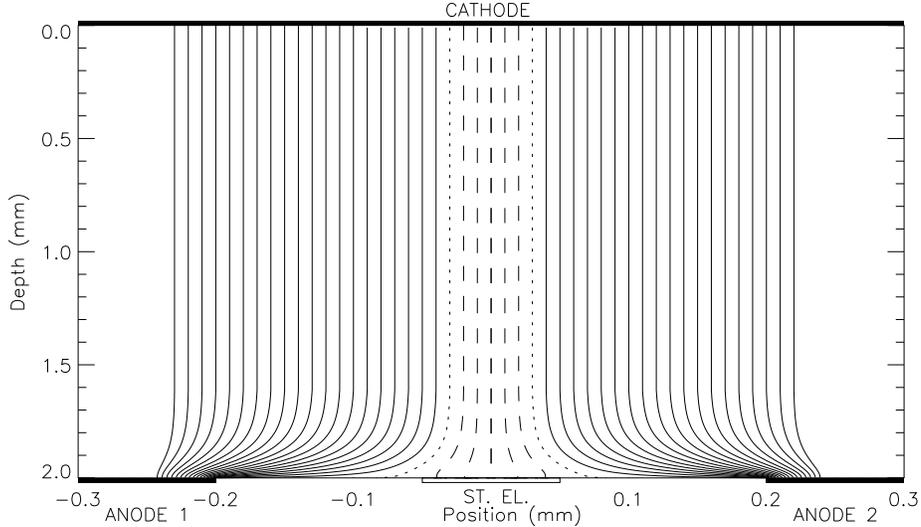}}
\caption[tra]
{\label{fig:tra}
Model calculation of electric field lines. Position and depth axes have
different scales. Field lines terminate at the anodes (solid lines), \SE s 
(dashed lines), or in the inter-electrode gaps (dotted lines).}
\end{figure}

  We developed a computer simulation of solid state detectors that predicts 
induced charge on each electrode for various electrode geometries and various 
interaction positions. We first calculate the electric fields and weighting 
potentials by using a general purpose program Maxwell 2D Field Simulator
\cite{MAXWELL}. We specify the appropriate geometry, electrode pattern and 
potentials. We use a dielectric constant of 10. The Maxwell software iteratively 
calculates the electrostatic field solution by finite element analysis. 
Fig.~\ref{fig:tra} shows the result of the Maxwell electric field calculation 
for the UCSD-WU strip detector described in Section 2. We then transport
the charges using the electric fields, electron and hole mobilities and trapping 
times and calculate the induced charge on each electrode using Ramo's weighting 
potential method\cite{Ramo39}. The simulation takes charge trapping into account,
but  the effects of space charge and detrapping are not included. (Further 
information on the model is presented by Kalemci et al\cite{Kalemci99}.)

 Our previous ``non-diffusive'' work modeled the ideal 
case where each interaction occurred at a point, with no lateral extension of 
the charge cloud. In this case the electrons and holes follow a single electric 
field line. Depending on the interaction position there are three cases. With 
reference to Fig.~\ref{fig:tra}, electrons can follow: (1) solid lines (89\% of 
the detector volume) and be collected by the anodes, (2) dashed lines (8\%) and 
be collected by the \SE, or, (3) dotted lines (3\%) and drift to the gap, in 
which case the signal is shared between an anode and the \SE.

  In Kalemci et al.\cite{Kalemci99} we discussed each case and showed 
experimental evidence that electric field lines end at the \SE\ and the gaps as
predicted by the simulation. Electrons reaching these positions will yield 
reduced anode signal, hence a low energy tail in the spectrum.

  However, this previous simulation did not include important additional 
effects which broaden the distribution of charges. These are described in the 
next section. By including these effects, we have obtained a more accurate 
model of our detector.

\section{Size of the electron cloud}

  When the incident X-ray interacts with the detector material, a \PE\ is 
ejected. This \PE\ loses energy by ionization, creating electron-hole pairs 
along its path until it is stopped. Therefore, X-ray interactions produce 
electron and hole distributions that are elongated along the path of the \PE. 
We call these distributions ``clouds''. The range of the \PE\ depends on the 
energy of the interaction. For example, a 40 keV \PE\ is stopped in 
$\rm 10\mu m$, whereas a 100 keV \PE\ is stopped in $\rm 47\mu m$\cite{NIST}. 
These clouds are not uniform in charge density, since electron-hole 
production increases towards the end of the track. For a 100 keV photoelectron,
 30\% of the electron-hole pairs are created in the last $\rm 7\mu m$ of its 
track. A K X-ray of $\rm \sim$25 keV may also be produced in the 
initial interaction with mean free path of $\rm \sim$$\rm 85\,\mu m$, and its 
charge cloud will add to the overall distribution of charge in the detector.

  Moreover, the clouds diffuse while drifting to the electrodes, making them 
larger. The \D\ of a material with concentration M is characterized by Fick's 
equation\cite{Beam65}:
\begin{equation}
\label{eq:Fick}
\rm D\,{\nabla^{2}M} = {\partial M\over{\partial t}}
\end{equation}  
where D is the diffusivity (or Einstein Coefficient). D can be obtained from
 the Einstein Relation, $\rm D=\mu kT/e$, and it is $\rm\sim$$\rm 26\,cm^{2}/s$
 for electrons at room temperature with mobility, $\rm \mu$, of
$\rm 1000\,cm^{2}/V\,s$.  For electrons, the travel time of $\rm \sim$
$\rm 0.2\,\mu s$ for a 2 mm thick detector is much shorter than the trapping 
time, $\rm \sim$$\rm 3\,\mu s$; therefore, we must use the time dependent 
solution. The one dimensional solution for concentration M(x,t) at position x 
and time t for a delta function initial concentration $\rm M_{o}\,\delta 
(x_{o},0)$ at position $\rm x_{o}$ is\cite{Beam65}
\begin{equation}
\label{eq:dif}
\rm M(x,t)={M_{o}\over{2\,(\pi\,D\,t)^{1/2}}}\;\;exp[\,{-(x-x_{o})^{2}\over{4\,D\,t}}\,]
\end{equation}
where t is  the time since the interaction. 
\begin{figure}[tb]
\centerline{\psfig{figure=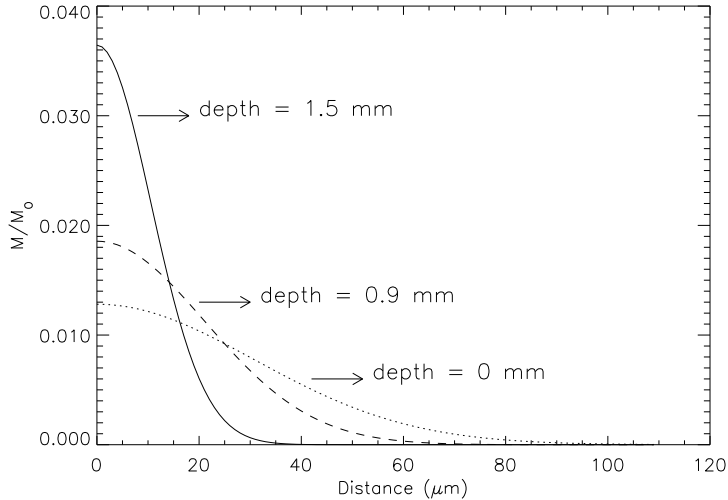,height=7cm}}
\caption[diff1]
{\label{fig:diff1}
Concentration of particles as a function of position for different interaction 
depths. Profiles show the concentration when the electrons have drifted to a 
depth of 1.7 mm. $\rm M_{o}$ is the initial concentration.}
\end{figure}

In general, at depths greater than $\rm\sim$1.7 mm, the lateral electric field
 is strong enough to compress the charge cloud and guide it to an anode (See 
Fig.~\ref{fig:tra}). Therefore, we ignore diffusion once the charge drifts to
depths greater than 1.7 mm. If $\rm t_{1.7}$ is the time elapsed between the 
interaction time and the time that the cloud reaches a depth of 1.7 mm, then
the concentration at this depth as a function of position can be obtained by 
substituting $\rm t_{1.7}$ in place of t in Equation 2. This time can be obtained
using our electric field calculations as follows:
\begin{equation}
\label{eq:t17} 
\rm t_{1.7}={{(0.17 - DOI)}\over{\mu\;E}}
\end{equation}
where DOI is the depth of interaction in cm, and E is the electric field component
perpendicular to the detector surface. This field is constant to 1\% between the
top of the detector and 1.7 mm, and has a value of $\rm \sim 930\;V\,/\,cm$. Since
$\rm t_{1.7}$ decreases with depth of interaction, the size of the cloud decreases
as depth of interaction increases (Fig.~\ref{fig:diff1}). 
\section{Measurements}

We studied the detector's response using collimated gamma-rays from a \CO\ 
radioactive source whose dominant emission is at 122 keV, with two weaker 
lines at 136 keV and 14.4 keV. The collimator uses a stack of precision, laser 
machined tantalum disks with holes ranging from 30 to $\rm 60\,\mu m$ to form 
a 3 mm thick layer with a fine hole passing through it\cite{SPIE2859}. The hole
 is tapered to give a uniform intensity in its $\rm 30\,\mu m$ aperture.
This  was placed $\rm \sim$0.5 mm from the detector, and the radioactive 
source was $\rm \sim$10 cm from the aperture. The position of the collimator 
was controlled by an X-Y stage with $\rm 10\,\mu m$ precision. 

  Two adjacent anodes, the steering electrode and a cathode were coupled 
to Amptek A250 charge sensitive preamplifiers whose signals were processed by 
shaping, amplifying and triggering circuits, and digitized by ADC's to give 
the total signal on each electrode. An on-line program processed the data and 
 built multiple spectra according to various event selection 
criteria\cite{SPIE3445,Slavis98}. If the signal on any electrode exceeded its 
threshold, the system was triggered and all pulse heights were measured and 
recorded to build event lists for off-line analysis.

  We scanned perpendicular to the anode strips with the $\rm 30\,\mu m$ 
collimator. The position with an equal number of counts at each anode was 
assumed to correspond to the center of the \SE. (See Section 6.2 for a 
discussion of this assumption.) Other positions were measured 
with respect to this point. The illumination point was centered on a cathode 
strip. Since anodes and cathodes are orthogonal, scanning the collimator 
across the anodes kept it centered on the cathode.

\begin{figure}[tb]
\centerline{\psfig{figure=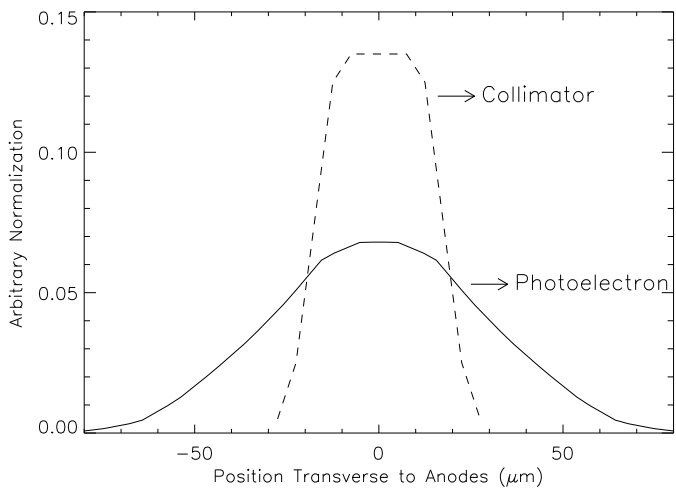,height=7cm}}
\caption[resp]
{\label{fig:resp}
The dashed line (Collimator) shows the lateral distribution of interaction 
positions with respect to collimator position. The solid line (Photoelectron) 
corresponds to the lateral distribution of endpoints of \PE\ 
tracks. }
\end{figure}

Interpretation of the measurements requires  knowledge of the distribution of 
electron clouds for a given collimator position. This was determined in two 
steps. First, we calculated the flux of photons from the collimator to obtain 
the lateral distribution of interaction positions. Then, we convolved this 
distribution with the \PE\ path distribution. Fig.~\ref{fig:resp} shows the 
distribution of end points of \PE\ tracks along with the distribution of 
initial positions of interactions. 

Since 122 keV gamma-rays have a range of $\rm \sim$1.5 mm, they interact 
throughout the detector, allowing us to study the dependence of charge sharing
on depth. Because of hole trapping, the cathode signal decreases as depth of 
interaction increases, and the ratio of cathode signals to anode signals 
indicates the interaction depth\cite{Kalemci99,He00}. We sorted events into 
depth ranges according to their cathode to anode ratios. Then, using the number
 of counts in each bin and the photoelectric absorption coefficient of CdZnTe 
for 122 keV photons, we calculated the boundaries of the depth ranges. 
\section{Results and interpretations}
\begin{figure}[tb]
\centerline{\psfig{figure=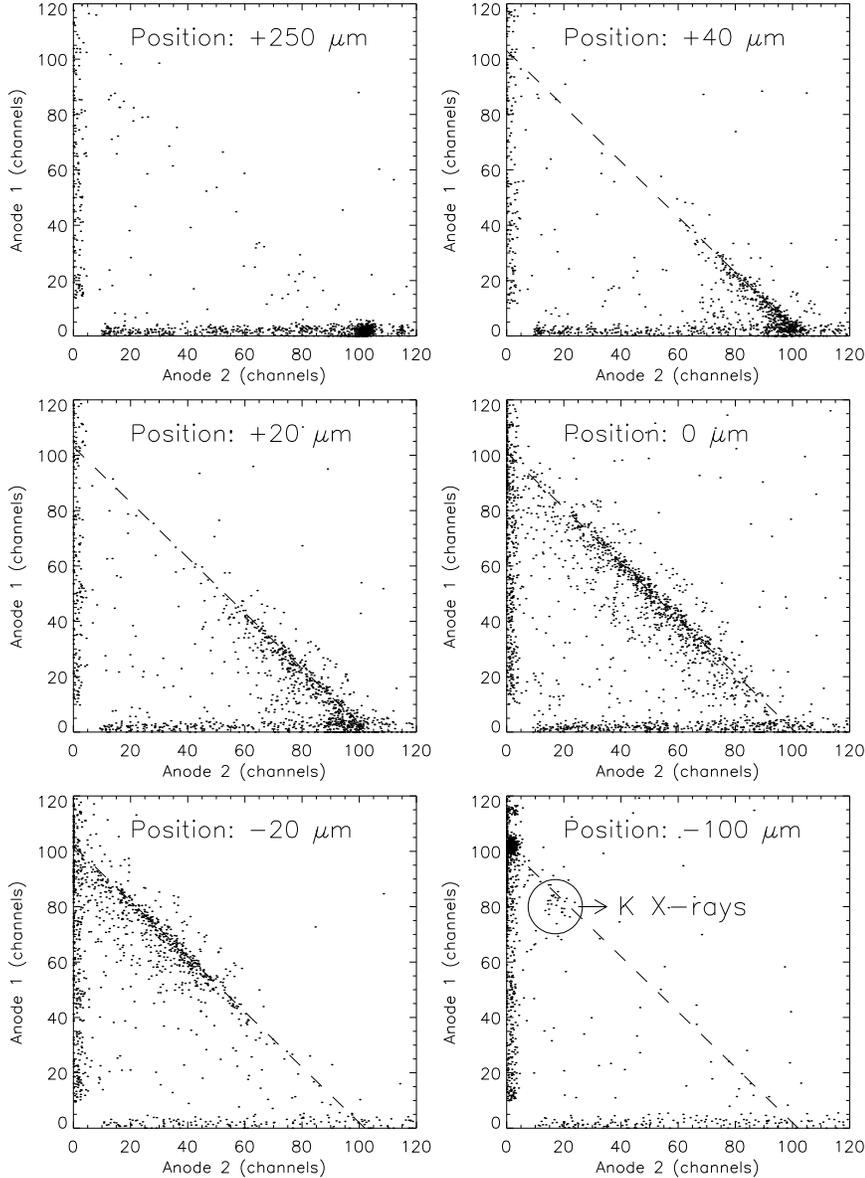}}
\caption[char_se]
{\label{fig:char_se}
Scatter plots of anode signals read out from the ADC. Each point represents 
one event. The positions represent the distance of the collimator from the 
center of the \SE\ (which is denoted by Position: $\rm 0\,\mu m$). The dashed 
lines represent the ideal case where sum of the charge collected by anode 1 
and anode 2 is proportional to the total charge deposited by 122 keV gamma-ray.
 $\rm 100\,\mu m$ away from the \SE, K X-rays appear as a separate feature, 
see text for more details. }
\end{figure}
\begin{figure}[b]
\centerline{\psfig{figure=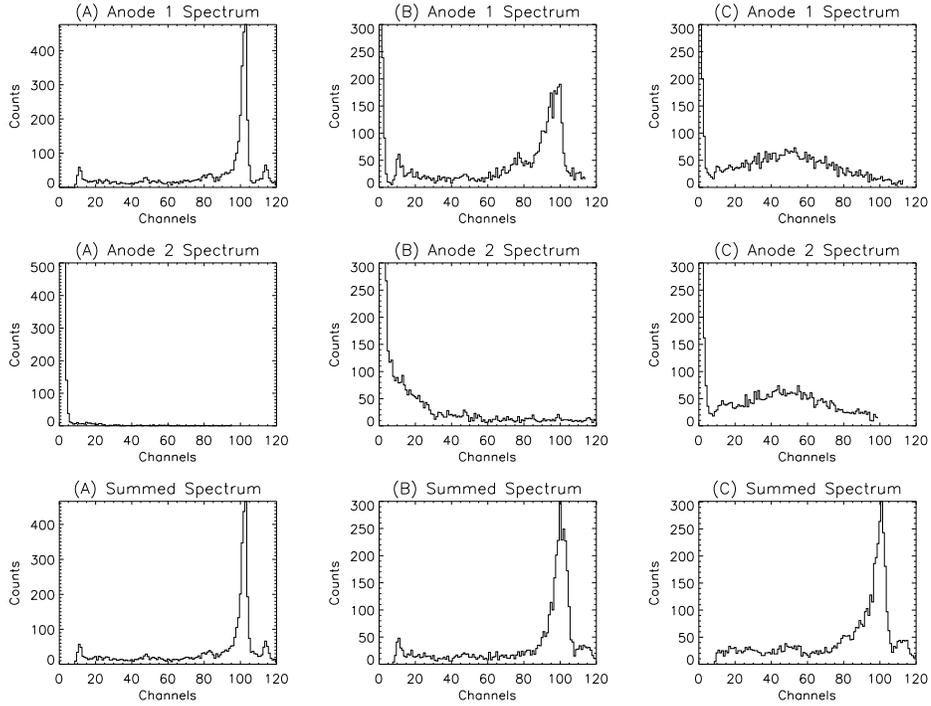,height=9.5cm}}
\caption[co_3dep]
{\label{fig:co_3dep}
Individual anode spectra and the summed anode spectra of \CO\ for the three 
different collimator positions (A, B, C) indicated in Fig.~\ref{fig:co_ps}}
\end{figure}
\begin{figure}[tb]
\centerline{\psfig{figure=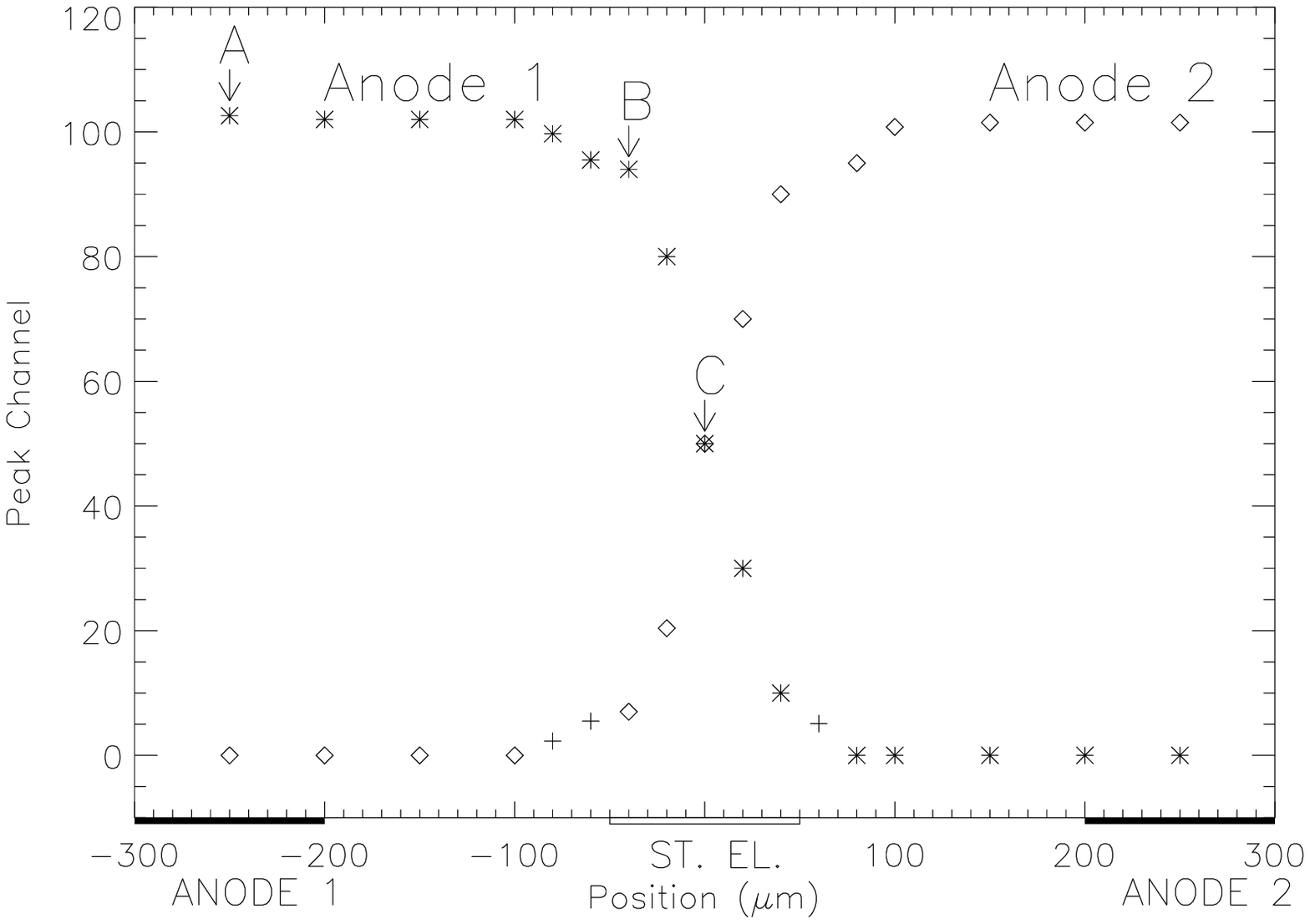,height=7cm}}
\caption[co_ps]
{\label{fig:co_ps}
Peak channels vs collimator positions for anode 1 and 2. The inferred values 
(see text) are shown with ``+'' symbols.}
\end{figure}
\subsection{Charge sharing}
For 122 keV events, the relative signals on the two neighbor anodes were found
to be a strong function of beam position, as shown in Fig.~\ref{fig:char_se}. 
Well away from the \SE, at $\rm +250\,\mu m$, there is negligible charge 
sharing. All 122 keV (channel $\rm\sim$100) and 136 keV (channel $\rm\sim$113)
 interactions are collected at anode 2. The remaining continuum is background, 
except at channel $\rm\sim$12 (14.4 keV)  where another line of \CO\ falls.

The Position +$\rm 250\,\mu m$ plot in Fig.~\ref{fig:char_se} is representative
of the results from all positions greater than $\rm 100\,\mu m$, where charge 
sharing occurs for less than 1\% of all interactions. For collimator positions 
less than $\rm 100\,\mu m$ from the \SE, some events fall on a diagonal line 
representing full energy in the summed anode signals (x+y=122 keV). At position 
$\rm 0\,\mu m$, three such lines are apparent, which correspond to 136 keV, 122 
keV  and $\rm \sim$97 keV (escape peak). Negative positions mean the collimator 
is near anode 1, so most charge is collected by it, as indicated in the 
$\rm -20\,\mu m$ and $\rm -100\,\mu m$ panels. Since the points lie on a diagonal
 line, little charge is lost to the \SE\ or the gap.

At $\rm 100\,\mu m$, there is a cluster of points circled in 
Fig.~\ref{fig:char_se} which deposit $\rm \sim 25\,keV$ (channel 20) in the 
neighbor anode. Since the K X-ray has an energy of $\rm \sim 25$ keV and has a 
mean free path of $\rm 85\,\mu m$, we interpret these points as events whose 
K X-rays have propagated beyond the center of the \SE\ so that they have been
 collected by the neighbor anode. 

We examined the same data in more detail using spectra (Fig.~\ref{fig:co_3dep})
for the three collimator positions shown in Fig.~\ref{fig:co_ps}. Position (A),
at $\rm -250\,\mu m$, was directly above anode 1, and all of the charge was 
collected by anode 1. The energy resolution of the summed spectrum is  3\% 
FWHM at 122 keV and the 14.4 keV line is  resolved. Position (B) was 
$\rm 40\,\mu m$ away from the center of the \SE\ and charge sharing clearly 
affects the spectra. The anode 1 spectrum has a low energy tail along with an 
escape peak at channel $\rm \sim 80$ while anode 2 has significant counts   
below channel 30 (36 keV). Compared to anode 1, the summed spectrum has a 
narrower peak and a less low energy tail. Position (C) was above the center of 
the \SE, and the two anodes have the same spectra, with peaks at channel 
$\rm\sim$50 (61 keV). In (C) all events shared charge, but a good spectrum was 
recovered by summing anode signals. 

Positions (B) and (C)'s summed spectra have broader peaks than that for 
Position (A). This is due to three effects. First, adding signals means adding 
the electronic noise in quadrature. Second, we only summed anodes when the
neighbor signal was larger than the noise level (4 channels). Therefore, if 
charge sharing occurred so that the signal on one anode was below the detection
threshold of the electronics, it was not summed, producing an artificial low 
energy tail. However, this happens only at a very specific distance from the 
\SE\, and possibly affected summed spectrum in position (B). Third, and most 
important, was incomplete charge collection due to collection of electrons by 
the \SE\ and the gap. This will be discussed in detail in Section 6.3.

Spectral peaks were measured for both anodes at 17 beam positions and the 
results are plotted in Fig.~\ref{fig:co_ps}. At three positions, the smaller 
peaks were too close to the noise to be determined. For these cases an 
``inferred peak'', equal to channel 102 minus the larger peak, is plotted. 
There is a smooth transition in peak channel from channel 102 (full signal) to
 zero signal over a $\rm \pm 80\,\mu m$ region around the \SE. Note that this 
result is the convolution of the intrinsic transition with the collimator 
response of $\rm \sim \pm 20\,\mu m$ (dashed line in Fig.~\ref{fig:resp}). 
Thus, the intrinsic transition occurs in $\rm \sim \pm 60\,\mu m$.

\subsection{Effects of diffusion and photoelectron range}

\begin{figure}[tb]
\centerline{\psfig{figure=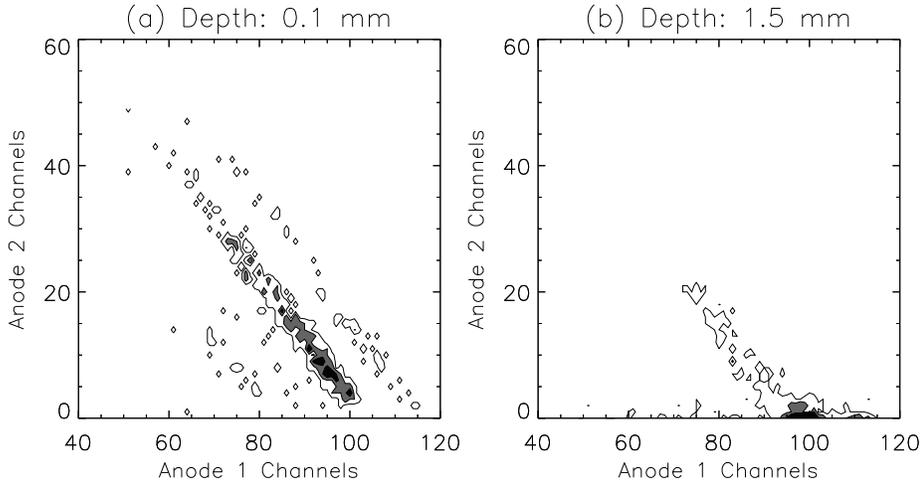}}
\caption[depth1]
{\label{fig:depth1}
The depth dependence of charge sharing. Figures are contour plots of 2D 
histograms of Anode 1 vs. Anode 2 signals. Darker regions indicate higher 
density of points. The collimator is at $\rm -40\,\mu m$. Panels (a) and (b) 
 correspond to average depths of 0.1 mm and 1.5 mm, respectively.  }
\end{figure}

We expect charge sharing to be greater for interactions near the cathode, since
 the cloud must drift through a greater depth, diffusing more. We showed this 
to be the case by comparing the sharing effects at the top and 
the bottom of the detector for a collimator position $\rm 40\,\mu m$ away from
 the center of the \SE, as shown in Fig.~\ref{fig:depth1}. Panel (a) is for 
interactions in  the top 0.2 mm of the detector and significant charge sharing 
occurs. However, most of the interactions at the bottom of the detector  are 
fully collected at anode 1, as shown in panel (b). 

Quantitative studies of diffusion were made by comparing the onset of 
sharing versus beam position with the predicted onset. This was calculated 
from the \PE\ range distribution (Fig.~\ref{fig:resp}) modified by the 
predicted broadening due to diffusion (Eq.~\ref{eq:dif} and 
Fig.~\ref{fig:diff1}). This produced a predicted charge distribution as a 
function of distance from the collimator and depth of interaction. We defined the
 onset of charge sharing for each depth as the collimator position at which the 
neighbor anode received more than 5\% of the signal for at least 5\% of all the 
events. For 5\% of all interactions, the \PE\ range is greater than 
$\rm 45\,\mu m$, as determined by integrating the solid curve in  
Fig.~\ref{fig:resp}. By using Eq.~\ref{eq:dif} and Eq.~\ref{eq:t17}, we predicted
 the distance beyond which 5\% of the electrons reside when the electron cloud 
reaches at a depth of 1.7 mm, and called it L(5\%). Then the sharing onset is 
expected to occur at $\rm 45\,\mu m\,+\,L(5\%)$. This distance was calculated for 
three depths and compared with measurements, with the results shown in 
Table.~\ref{table:qud}.
\\ 
\begin{table} [h]   
\caption[qud]
{\label{table:qud}
Predicted and observed onset of charge sharing}
\begin{center}       
\begin{tabular}{|c|c|c|c|c|c|} 
\hline
\rule[-1ex]{0pt}{3.5ex} Avg. Depth & Depth Range& L(5\%) & Predicted  & 
Observed & No sharing \\
\hline
\rule[-1ex]{0pt}{3.5ex}  0.1 mm & 0 - 0.2 mm & $\rm 50_{-0}^{+1}\,\mu m$ & 
$\rm 95_{-10}^{+1}\,\mu m$ & 
$\rm 95\pm\,5\,\mu m$  & $\rm 105\pm\,5\,\mu m$ \\
\hline
\rule[-1ex]{0pt}{3.5ex}  0.44 mm & 0.35 - 0.53 mm & $\rm 43_{-0}^{+2}\,\mu m
$ & $\rm 88_{-10}^{+2}\,\mu m$ & $\rm 80\pm\,5\,\mu m$ & $\rm 100\pm\,5\,\mu m$
 \\
\hline
\rule[-1ex]{0pt}{3.5ex}  1.1 mm & 0.9 - 1.3 mm & $\rm 30_{-0}^{+4}\,\mu m$ & 
$\rm 75_{-10}^{+4}\,\mu m$ & $\rm 65\pm\,5\,\mu m$ & $\rm 85\pm\,5\,\mu m
$  \\
\hline
\end{tabular}
\end{center}

 \end{table}
 
Average depths, depth ranges, L(5\%) values, and predicted onsets are shown in 
the first four columns of Table 1. In the last two columns are the collimator 
positions where the onsets were observed and the nearest position where 
no sharing is observed. The results show that observational onsets are within
the uncertainties of the estimated onsets.  

Uncertainties in the observations correspond to the positioning of the collimator,
 and uncertainties in the calculations were combinations of various effects such 
as sampling preferentially lower depths within one depth range, uncertainty of 
charge sharing due to K X-ray interactions, and the assumption that the initial 
concentration of the electron cloud is a delta function. We also assumed that
the point where charge is shared equally is the center of the \SE. We did not 
have an experimental measurement to verify this assumption; however, we observed
that the charge is shared equally at each depth range for this collimator 
position. Even if this position does not correspond to the center of the \SE\, 
it still defines the plane beyond which the charges are collected by the neighbor
anode. Since the collimator positions are measured with respect to this plane, 
this assumption does not introduce  uncertainties in our results. We also 
estimated the contribution of holes to the charge sharing phenomenon using our 
simulations described in Kalemci et al \cite{Kalemci99}. We used a hole mobility 
of $\rm 40\,cm^{2}/V\,s$ and a hole trapping time of 650 ns in our simulations. 
The results showed that at the depth range the sharing onsets were calculated, 
0.1-1.3 mm, the holes contribute less than 1\% to the neighbor anode signals and 
is neglected.

\subsection{Charge loss to the steering electrode and gaps}

As noted earlier, for interactions within $\rm \sim$$\rm 60\,\mu m$ of the 
\SE, the summed anode signal is reduced by a few percent and shows tailing. A 
simulation neglecting \D\ predicts that electrons from these events would drift
 only to the \SE s, producing no signal on the anodes. We showed in Kalemci et 
al.\cite{Kalemci99} that, for some of the interactions, the \SE\ has positive 
signals, which means that it collects electrons for some events. However, if 
the number of field lines ending at the \SE\ is smaller than predicted by the 
model and/or the electron cloud is larger than $\rm \sim$$\rm 100\,\mu m$, only
 a small fraction of the electrons will be collected by it. This could explain 
the lack of full energy signals on the \SE.

It is difficult to interpret the \SE\ pulses to study such effects. Since \SE s
are joined together their weighting potentials extends throughout the detector.
(See Kalemci et al.\cite{Kalemci99} for weighting potential calculations.) They 
are sensitive to both electron and hole transport at all depths of interaction
and one can not extract the signal solely due to electrons reaching the \SE.
Therefore, to look for a charge loss to the \SE\ or the gaps, we analyzed summed 
anode spectra, since anode signals are not as depth dependent as \SE\ signals 
due to the small pixel effect. 

\begin{figure}[tb]
\centerline{\psfig{figure=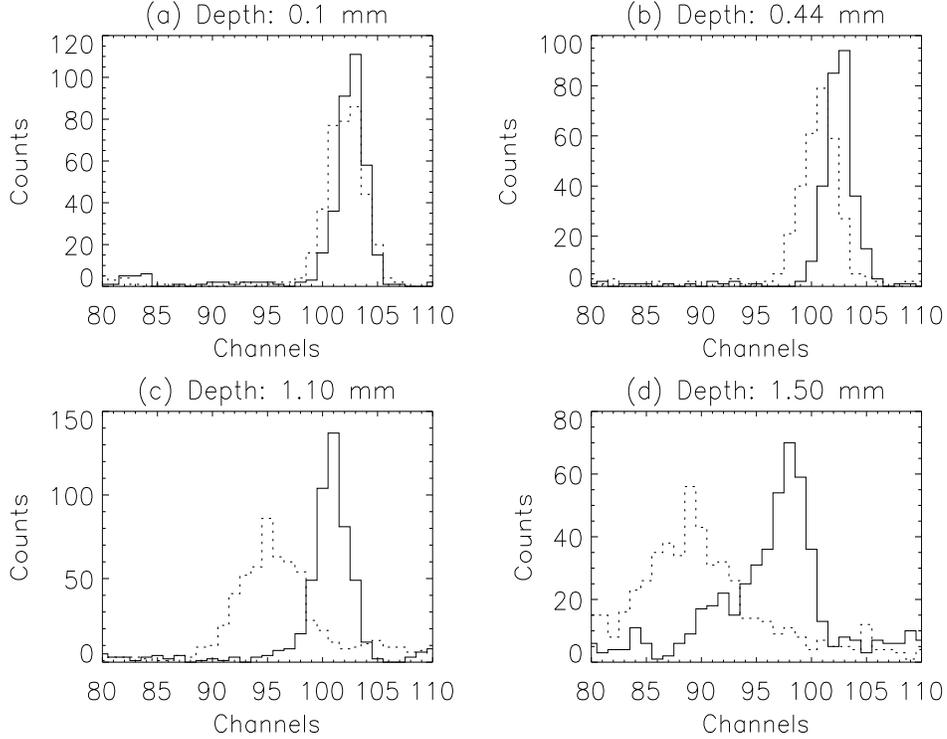,width=13cm,height=10cm}}
\caption[dif_def]
{\label{fig:dif_def}
Depth dependence of signal lost to the \SE. Each panel contains two spectra,
one for collimator directly above one anode (solid line) and other directly 
above the \SE. These show that, as the interactions get deeper, the \SE\ 
collects more charge, reducing the summed anode signal.}
\end{figure}

We compared the summed anode spectrum for a collimator position above anode 2 
with that of above the \SE. This analysis was done for various depth ranges, and 
the results are shown in Fig.~\ref{fig:dif_def}. At each depth, we considered the 
shift of line centroid over the \SE\ relative to that over the anode. Near the 
top of the detector (0.1 mm), the shift is $\rm \sim$1\%. For deeper interactions,
 the shift became larger. For the average depth of 1.5 mm, the centroid shifted 
$\rm\sim$9\%. The increase of centroid shift with increasing depth is 
consistent with the \D\ of electron cloud with time, because, for deeper 
interactions, the electrons have less time to diffuse and the charge cloud is 
smaller. Therefore, the \SE\ collects  more of the electrons.

These data allow us to estimate the size of the detector region where the field
 lines end at the \SE. This is equal to the lateral extent of the part of the 
charge cloud which drifts to the \SE. At the depth range, 1.1 mm to 1.7 mm, the
maximum charge lost is $\rm\sim$15\% (Fig 10d). Assuming these events occurred at 
1.7 mm, and the \PE s moved towards the \SE\ with no additional broadening, a 
diffusion time of t=40 ns results in 15\% of the electrons being within 
$\rm \pm 2\,\mu m$ of the charge cloud's center. Thus, the region with field
lines ending on the \SE\ measures just $\rm \pm 2\,\mu m$. However, our 
electric field calculation (Fig.~\ref{fig:tra}) predicted a $\rm \pm 20\,\mu m$ 
region. This would result in the majority of electrons being collected by the 
\SE\ for interactions above it at any depth. We therefore conclude that the model 
overestimates the number of field lines ending at the \SE.

\section{Conclusion}

We have investigated charge collection by various electrodes and charge
sharing between two anodes of a 2 mm thick CZT strip detector with 
$\rm 500\,\mu m$ pitch electrodes. A $\rm 30\,\mu m$ collimated beam of 122 
keV gamma-rays was used to study the detector response at various positions 
and the cathode signal was used to infer depth of interaction. The results 
were interpreted in the context of our charge collection model.  

We found that \D\ is a very important mechanism causing charge sharing. For an
interaction with minimal depth in the detector, the electron cloud size 
increases to $\rm\sim$$\rm 100\,\mu m$  due to diffusion while it drifts 2 mm 
to the anodes in a $\rm\sim$1000 V/cm field.  Charge sharing occurs over 24\% 
of our detector volume for our specific electrode pattern. 

The solution of Fick's equation predicts that the size of the cloud changes as
a function of depth. We tested this by mapping the onset of charge sharing as 
a function of interaction depth. The results were in good agreement with the
calculated onsets. Another important mechanism affecting charge sharing is the
 photoelectron range. Without considering its effects, the predicted onsets in 
Table.~\ref{table:qud} would be reduced by $\rm\sim$$\rm 20\,\mu m$, contrary 
to the observed onsets.

Our detector model predicts that interactions above the steering electrode 
would have the majority of their electrons collected by the \SE, but the 
laboratory tests indicate that this is not the case. Rather, $\rm >$90\%
of the electrons reach the anodes for interactions only 0.5 mm above the \SE. 
This implies that the simulation overestimates the number of field lines ending
 at the \SE. A similar effect was reported in a study by Prettyman et 
al.\cite{Prettyman99} where they discussed possible causes, such as 
band bending near the anodes, surface effects, and microscopic defects.

In the region where charge sharing occurred, the summed anode signal was at 
91\% to 99\% of that for single anode events. The deficit represents electrons
 lost to the \SE\ and the gaps. It appears to be possible to correct for
these effects. First, the ratio of cathode-to-anode signals can be used to 
infer the depth, allowing a depth dependent correction to the energy for hole 
trapping as shown by Kalemci et al.\cite{Kalemci99}. Second, for shared events,
 the ratio of the signals on the two anodes indicates the transverse 
position which can be used, along with the depth, to correct for the signal 
lost to the \SE\ and the gaps. Such a full correction is the subject of future 
work.
%
\ack{This work was supported by NASA SR\&T Grant NAG5-5111 and NASA Grant 
NAG5-8498. Technical support was provided at UCSD by Fred Duttweiler, George 
Huszar, Charles James, Phillippe Leblanc, David Malmberg, Ron Quillan and Ed 
Stephan. Authors would like to thank Richard Rothschild, William Heindl, John 
Tomsick and R. T. Skelton for their valuable discussions. We also would like to 
thank Kimberly Slavis at WU for her contributions. Emrah Kalemci was partially 
supported by TUBITAK. }


%
   \end{document}